# A New Approach for 4DVar Data Assimilation

Xiangjun Tian[1,2*], Aiguo Dai[3], Xiaobing Feng[4], Hongqin Zhang[1,2], Rui Han[1,2] and Lu Zhang[1,2]

[1]ICCES, Institute of Atmospheric Physics, Chinese Academy of Sciences, Beijing, 100029, China

[2]University of Chinese Academy of Sciences, Beijing, 100049, China

[3]Department of Atmospheric and Environmental Sciences, University at Albany, SUNY,

1400 Washington Ave. Albany, NY 12222, U.S.A.

[4]Department of Mathematics, the University of Tennessee, Knoxville, TN 37996, U.S.A.

---

[*]Corresponding author e-mail: Xiangjun Tian, tianxj@mail.iap.ac.cn

**Four-dimensional variational data assimilation (4DVar) has become an increasingly important tool in data science with wide applications in many engineering and scientific fields such as geoscience[1-12], biology[13] and the financial industry[14]. The 4DVar seeks a solution that minimizes the departure from the background field and the mismatch between the forecast trajectory and the observations within an assimilation window. The current state-of-the-art 4DVar offers only two choices by using different forms of the forecast model: the strong- and weak-constrained 4DVar approaches[15-16]. The former ignores the model error and only corrects the initial condition error at the expense of reduced accuracy; while the latter accounts for both the initial and model errors and corrects them separately, which increases computational costs and uncertainty. To overcome these limitations, here we develop an integral correcting 4DVar (i4DVar) approach by treating all errors as a whole and correcting them simultaneously and indiscriminately. To achieve that, a novel exponentially decaying function is proposed to characterize the error evolution and correct it at each time step in the i4DVar. As a result, the i4DVar greatly enhances the capability of the strong-constrained 4DVar for correcting the model error while also overcomes the limitation of the weak-constrained 4DVar for being prohibitively expensive with added uncertainty. Numerical experiments with the Lorenz model show that the i4DVar significantly outperforms the existing 4DVar approaches. It has the potential to be applied in many scientific and engineering fields and industrial sectors in the big data era because of its ease of implementation and superior performance.**

Data assimilation (DA), whose latest development is represented by the four-dimensional variational (4DVar) approach[17-18], has experienced explosive growth and development, especially in the context of the big data era[1-14]. This is because DA can effectively incorporate the time series of observational data into model simulations and predictions to improve estimates of all the current and future states of a natural (e.g., the atmosphere)[1-13] or social (e.g., the financial markets) system[14]. For example, most numerical weather prediction (NWP) centers around the world have adopted the 4DVar approach[8-12] to assimilate asynchronous observations simultaneously, which has greatly improved the accuracy of weather prediction.

Currently, the 4DVar only has two approaches: the strong- and weak-constrained methods depending on whether the 4DVar solution is required to satisfy the forecast model exactly[15-16]. The strong-constrained 4DVar (s4DVar) assumes the forecast model is perfect and all errors in the prediction originate from the initial conditions. This is clearly unrealistic in many cases, and recent studies[15-16,19-21] show that incorporating the model error into the 4DVar improves its performance. The weak-constrained 4DVar (w4DVar) allows for both initial and model errors and it corrects them separately[15-16], which adds computational costs significantly and increases uncertainty. To reduce the computational costs, various simplifications are used to specify the model error in the w4DVar[20-21]. However, these simplifications still have many issues; for example, it is very difficult or expensive to determine the parameters used to represent the model error covariance[21].

To overcome the limitations of both the s4DVar and w4DVar, we develop a new 4DVar approach in which all errors are treated indistinctly and corrected simultaneously. Our approach is based on the observation that the influence of all errors would decay gradually in time if the correction-term is incorporated into the model integration sequentially.

Here, we first show a less recognized fact that the s4DVar actually has a hidden mechanism that can correct the model error at the initial (i.e., analysis) time. The s4DVar seeks an analysis increment $\mathbf{x}' \in \mathbb{R}^{m_x}$ of the initial background condition $\mathbf{x}_0$, such that $\mathbf{x}'$ minimizes the following cost function (so that $\mathbf{x}_0^* = \mathbf{x}_0 + \mathbf{x}'$ represents the corrected estimate of the state variable $\mathbf{x}$ at the initial time $t_0$, see Fig. 1)[20]:

$$\ell = \frac{1}{2}(\mathbf{x}_0')^T \mathbf{B}^{-1}(\mathbf{x}_0') + \frac{1}{2}\sum_{n=0}^{S}\left[\mathbf{h}_n(\mathbf{x}_{k_n}) - \mathbf{y}_{obs,n}\right]^T \mathbf{R}_n^{-1}\left[\mathbf{h}_n(\mathbf{x}_{k_n}) - \mathbf{y}_{obs,n}\right], \quad (1)$$

under the constraint $\mathbf{x}_k = \mathbf{f}_k(\mathbf{x}_{k-1})$, where $\mathbf{x}'$ is assumed to be Gaussian with the covariance matrix $\mathbf{B} \in \mathbb{R}^{m_x \times m_x}$, $\mathbf{x}_k$ is the state values at time step $k$, $k_n$ is the time step at the observational time level $t_n$ and $\mathbf{f}_k$ is the forecast model from time step $k-1$ to $k$. Other symbols follow the general convention and are described in Extended Data (ED) Table 1.

The traditional s4DVar assumes that the forecast model $\mathbf{x}_k = \mathbf{f}_k(\mathbf{x}_{k-1})$ describes the underlying system exactly[15], thus the model error is negligible. However, the model error is often non-negligible due to errors from discretization of continuous fields, approximation of certain physical processes, parameter uncertainties, boundary conditions, and round-off errors. Moreover, various studies show that the model error prevails over the initial error in many circumstances[22]. The use of the 4DVar at several major NWP centers also indicated that the 4DVar needs to account for model errors[8-12] because NWP models are imperfect.

The s4DVar can still deliver comparable performance with or even outperforms (ED Figs.1-2) one version of the w4DVar for a very special case with perfect initial conditions but random model errors. This case clearly does not fulfill the perfect model assumption, yet the s4DVar still worked reasonably well. This is because the s4DVar is formulated as a constrained optimization problem, which itself (eq. 1) does not require the forecast model $\mathbf{f}_k$ to be perfect without errors. From this view point, the analysis increment $\mathbf{x}'$ derived from the s4DVar is actually an integral "correction term" to correct all errors (including the initial and model errors) indiscriminately at the analysis time $t_0$. Thus, the widely-accepted perfect model assumption in the s4DVar is actually unnecessary for deriving the analysis increment $\mathbf{x}'$ as long as one considers the increment $\mathbf{x}'$ as for correcting both initial and model errors and allows $\mathbf{f}_k$ to contain errors. This important point seems have not been recognized in the DA community. However, the s4DVar focuses solely on the correction term at the initial analysis time without explicit corrections for the model errors at other times in the assimilation window; thus it is expected to perform poorly when the model error is large.

To address this deficiency, the recently-introduced w4DVar[15-16] seeks the analysis increment ($\mathbf{x}'$) of the initial conditions and the model-error correction terms $\boldsymbol{\varepsilon}_k^m \in \mathbb{R}^{m_x}$, such that $\mathbf{x}'$ and $\boldsymbol{\varepsilon}_k^m$ minimize the following cost function[15] (see Fig. 1):

$$\ell = \frac{1}{2}(\mathbf{x}')^T \mathbf{B}_0^{-1}(\mathbf{x}') + \frac{1}{2}\sum_{n=0}^{S}\left[\mathbf{h}_n(\mathbf{x}_{k_n}) - \mathbf{y}_{obs,n}\right]^T \mathbf{R}_n^{-1}\left[\mathbf{h}_n(\mathbf{x}_{k_n}) - \mathbf{y}_{obs,n}\right] + \frac{1}{2}\sum_{k=0}^{L_0}(\boldsymbol{\varepsilon}_k^m)^T \mathbf{Q}_k^{-1}(\boldsymbol{\varepsilon}_k^m), \quad (2)$$

under the state constraint $\mathbf{x}_k = \mathbf{f}_k(\mathbf{x}_{k-1}) + \boldsymbol{\varepsilon}_k^m$, where $\mathbf{x}'$ and $\boldsymbol{\varepsilon}_k^m$ are assumed to be Gaussian with the covariance matrices $\mathbf{B}_0 \in \mathbb{R}^{m_x \times m_x}$ and $\mathbf{Q}_k \in \mathbb{R}^{m_x \times m_x}$, respectively. Here $\mathbf{f}_k$ may also contain a model error; but different from the s4DVar, an explicit correction term $\boldsymbol{\varepsilon}_k^m$ is used to adjust the state $\mathbf{x}_k$ at all times. The size of the w4DVar optimization increases greatly because of the extra variables $\boldsymbol{\varepsilon}_k^m$; hence, its practical applications are very limited. The situation is further aggravated by our limited knowledge about the general form of the model error terms $(\boldsymbol{\varepsilon}_k^m)$[19-20].

Thus, the s4DVar handles all errors as an integral term but only at the analysis time, while the w4DVar attempts to differentiate the initial and model errors and to correct them separately, resulting a much more complicated and unpractical problem due to the high complexity of error sources. Here we propose a new approach, which takes the advantages of both the s4DVar and the w4DVar but avoiding their deficiencies: To extend the s4DVar's strategy of correcting the initial and model errors as a whole to other times in the assimilation window, we introduce an additive correction term $\mathbf{x}'_{k-1}$ into the state variable $\mathbf{x}_{k-1}$ at time step $k-1$ to obtain a corrected state variable $\mathbf{x}_{k-1}^* = \mathbf{x}_{k-1} + \mathbf{x}_{k-1}'$, which is then used to obtain the state variable at time step $k$: $\mathbf{x}_k = \mathbf{f}_k(\mathbf{x}_{k-1}^*)$. Moreover, we propose the following exponential function for the integral correction term $\mathbf{x}_k'$ based on a thorough analysis of the error evolution process (see Methods):

$$\mathbf{x}_k' = \begin{cases} \mathbf{x}'(1-\upsilon), & k = 0 \\ \mathbf{x}' \dfrac{1}{1-\upsilon}\left[\upsilon^2 + (1-2\upsilon)\upsilon^k\right], & 1 \leq k \leq L_0 \end{cases}. \tag{3}$$

Here $\mathbf{x}'$ is the maximum integral correction term and $\upsilon \in [0, 0.5]$ is a pre-determined parameter. Eq. (3) indicates that the integral correction term ($\mathbf{x}_k'$) decreases exponentially with time index $k$ (ED Fig.3) for $0 < \upsilon < 0.5$. Our new integral correcting 4DVar (i4DVar, Fig.1) method is defined by seeking the maximum correction error $\mathbf{x}'$ that minimizes the following cost function:

$$\ell = \frac{1}{2}(\mathbf{x}')^T \mathbf{Q}_b^{-1}(\mathbf{x}') + \frac{1}{2}\sum_{n=0}^{S}\left[\mathbf{h}_n(\mathbf{x}_{k_n}) - \mathbf{y}_{obs,n}\right]^T \mathbf{R}_n^{-1}\left[\mathbf{h}_n(\mathbf{x}_{k_n}) - \mathbf{y}_{obs,n}\right], \tag{4}$$

under the constraint (with the predetermined parameter $\upsilon$):

$$\mathbf{x}_k = \begin{cases} \mathbf{f}_k\left(\mathbf{x}_{k-1} + \mathbf{x}'(1-\upsilon)\right), & k = 1 \\ \mathbf{f}_k\left(\mathbf{x}_{k-1} + \mathbf{x}' \dfrac{1}{1-\upsilon}\left[\upsilon^2 + (1-2\upsilon)\upsilon^{k-1}\right]\right), & 2 \leq k \leq L_0 \end{cases}, \tag{5}$$

where $\mathbf{x}'$ is assumed to be Gaussian with the covariance matrix $\mathbf{Q}_b \in \mathbb{R}^{m_x \times m_x}$. Intuitively, the structure of $\mathbf{Q}_b$ should be similar to $\mathbf{B}$ of the s4DVar but with different standard deviations. The i4DVar degenerates to the s4DVar when $\upsilon = 0$ and to a special w4DVar case (i.e. with a constant bias error) when $\upsilon = 0.5$. Thus, the i4DVar can completely recover the s4DVar and partially recover the w4DVar. Compared with the s4DVar, the size of the i4DVar optimization is not changed and its implementation could be accomplished with a minimum added computational cost because that they differ only slightly in the way to integrate the forecast model (namely, $\mathbf{f}_k(\mathbf{x}_{k-1})$ vs. $\mathbf{f}_k(\mathbf{x}_{k-1} + \mathbf{x}_{k-1}')$) during the optimization process.

The i4DVar outperforms both the s4DVar and w4DVar methods in the framework of the Lorenz model[23] under scenarios with model errors from different sources, as it produces considerably smaller RMS errors (Fig.2) in the assimilation mode. Comparably, the disadvantage of the s4DVar becomes increasingly apparent towards the later part of the assimilation period due to the increased influence of the model errors. These results demonstrate that the exponential function used in the i4DVar represents well the evolution of a significant part of the total error in $\mathbf{x}_k$ under various model-error scenarios and leads to a large reduction of the RMS errors. Further, the i4DVar's performance is insensitive to different types of the model errors, and it performs almost as good as or even slightly outperforms the s4DVar for the perfect model case (ED Figs.4-5). In contrast, the explicit model-error representation in the w4DVar is ineffective, leading to its inferior performance than the s4DVar except for the last half of the assimilation period (Fig.2), when the tangent linear model of the forecast model was capable of characterizing the error evolution generally. Thus, it could be problematic to use an inaccurate form of the model error in the w4DVar.

In the forecast mode, we used $\mathbf{x}'_{L_0+1} = \mathbf{x}' \frac{1}{1-\upsilon}\left[\upsilon^2 + (1-2\upsilon)\upsilon^{L_0}\right]$ as an approximation of the model error to adjust the forecast at each time step, as the influence of the initial error is largely eliminated outside of the assimilation window. Again, the i4DVar significantly outperforms the other two approaches consistently (Fig. 3). This further demonstrates that the strategy of treating all errors as a whole is effective and the exponentially decaying function works.

In summary, we showed that by introducing a decaying error correction term for all the time steps in the assimilation window, it is possible to significantly improve the performance of the 4DVar using the Lorenz model. The proposed i4DVar approach, combined with the ensemble-based nonlinear least squares fast algorithm[24] (also see Methods) that does not require an adjoint model[17] for solving the minimization problem, can be easily applied to other DA cases with minimum added costs, and it has the potential to significantly advance DA.

**Supplementary Information** is available in the online version of the paper.

**Acknowledgments.**

The work of the first author was partially supported by the National Key Research and Development Program of China (2016YFA0600203), the National Natural Science Foundation of China (41575100) and the High-resolution Earth Observation System Major Special Project (CHEOS) (32-Y20A17-9001-15/17). A. Dai was supported by the U.S. National Science Foundation (Grant #AGS–1353740), the U.S. Department of Energy's Office of Science (Award #DE–SC0012602), and the U.S. National Oceanic and Atmospheric Administration (Award #NA15OAR4310086).

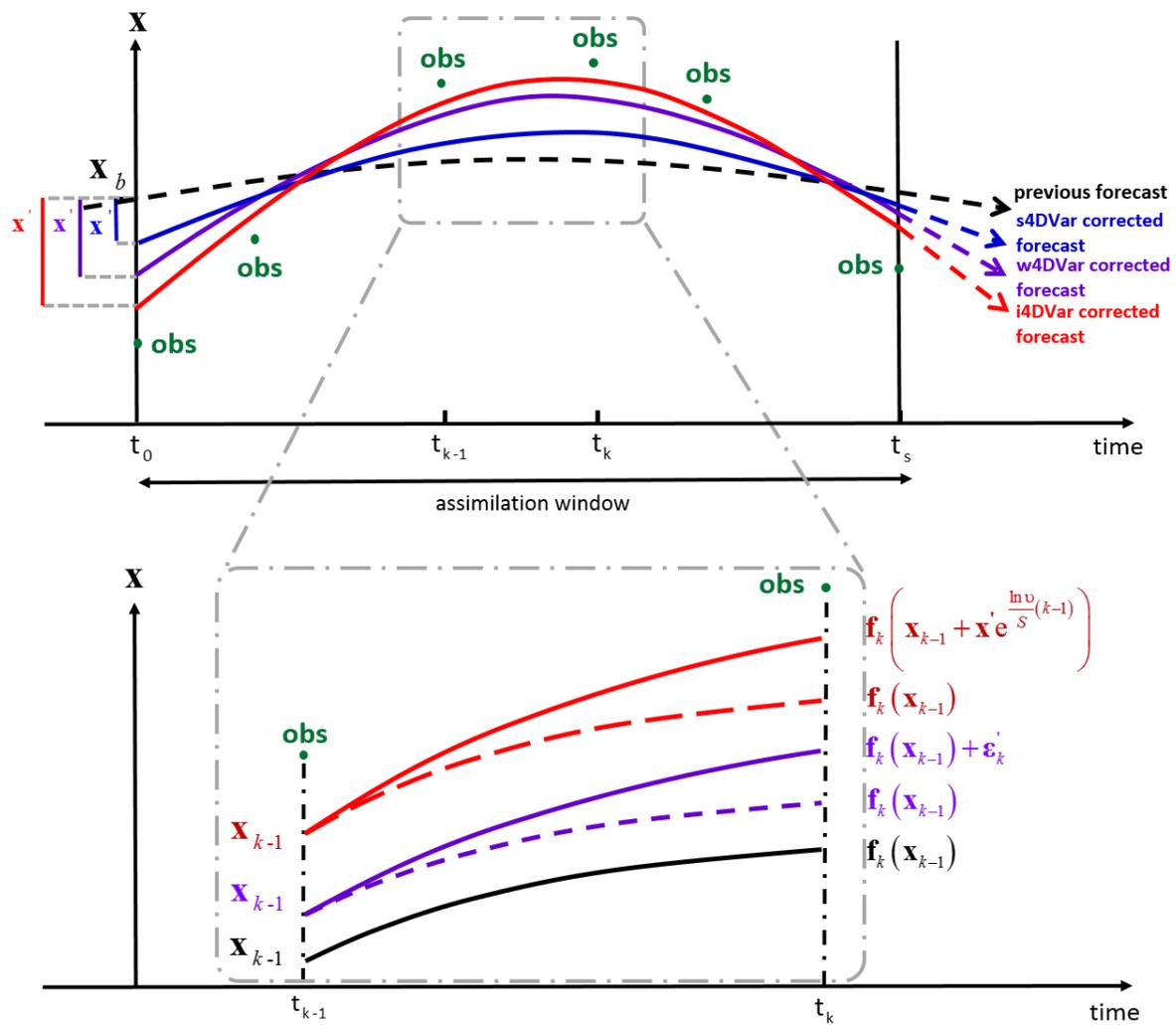

**Figure 1 | Schematic diagram illustrating the three 4DVar methods.** Over the assimilation window, each 4DVar is used to assimilate the observations (green dots) using a segment of the previous forecast as the background (or initial) state $\mathbf{x}_b$ (black dashed line) to yield its corresponding optimized fit to all the observations within the window with its own way to integrate the forecast model , as shown in the lower panel. The optimized fields [i.e., the sold lines determined by the analysis increment $\mathbf{x}'$ (blue) for s4DVar, the analysis increment $\mathbf{x}'$ and the model-error correction terms $\varepsilon_k^m$ for w4DVar (purple), and the maximum correction-term $\mathbf{x}'$ for i4DVar (red), respectively] are then used to improve the subsequent forecast.

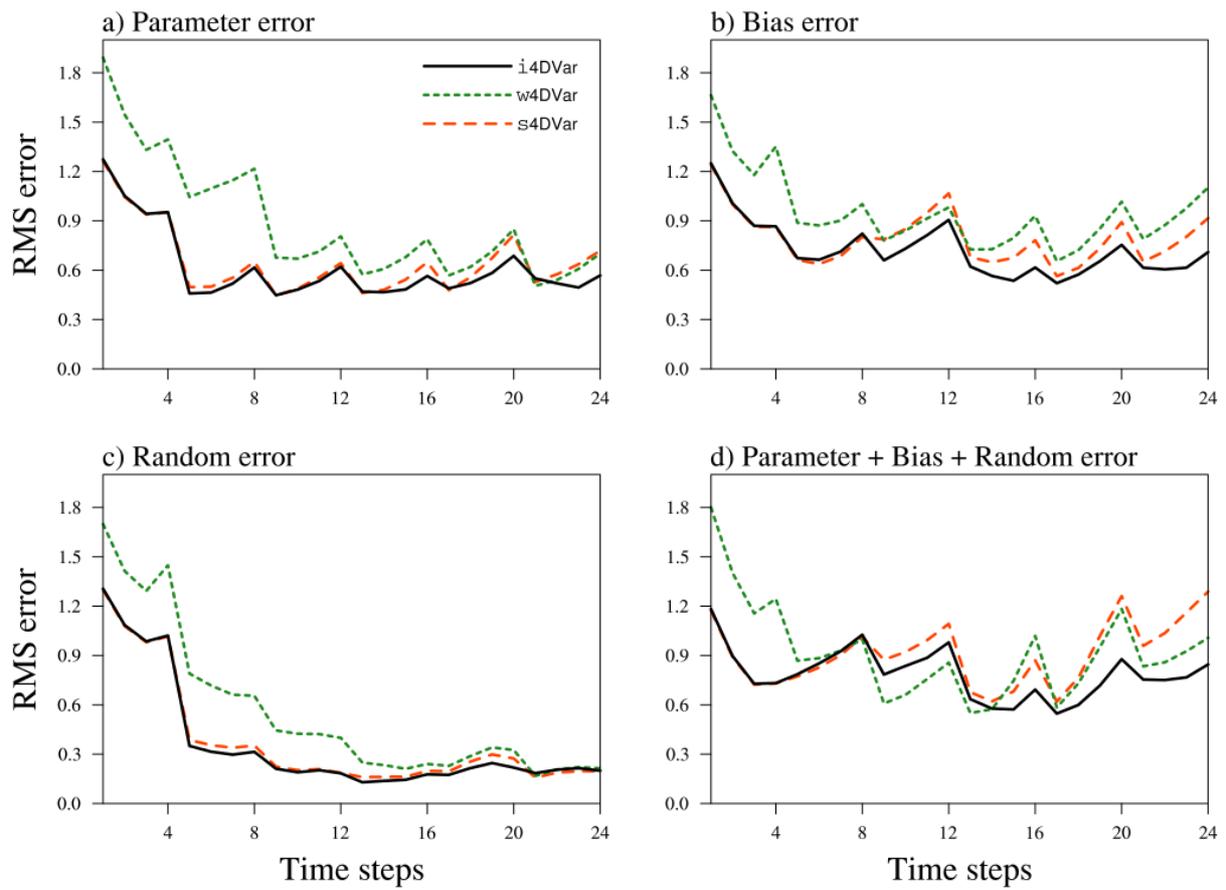

**Figure 2 | Time series of the 4DVar-assimilated root mean square (RMS) errors using three different methods. a**, The 24-step (6-window) assimilation results under the parameter error scenario. **b,** Same as **a**, but for the bias error scenario. **c**, Same as **a**, but for the random error scenario. **d**, Same as **a,** but for the parameter error + bias error + random error scenario.

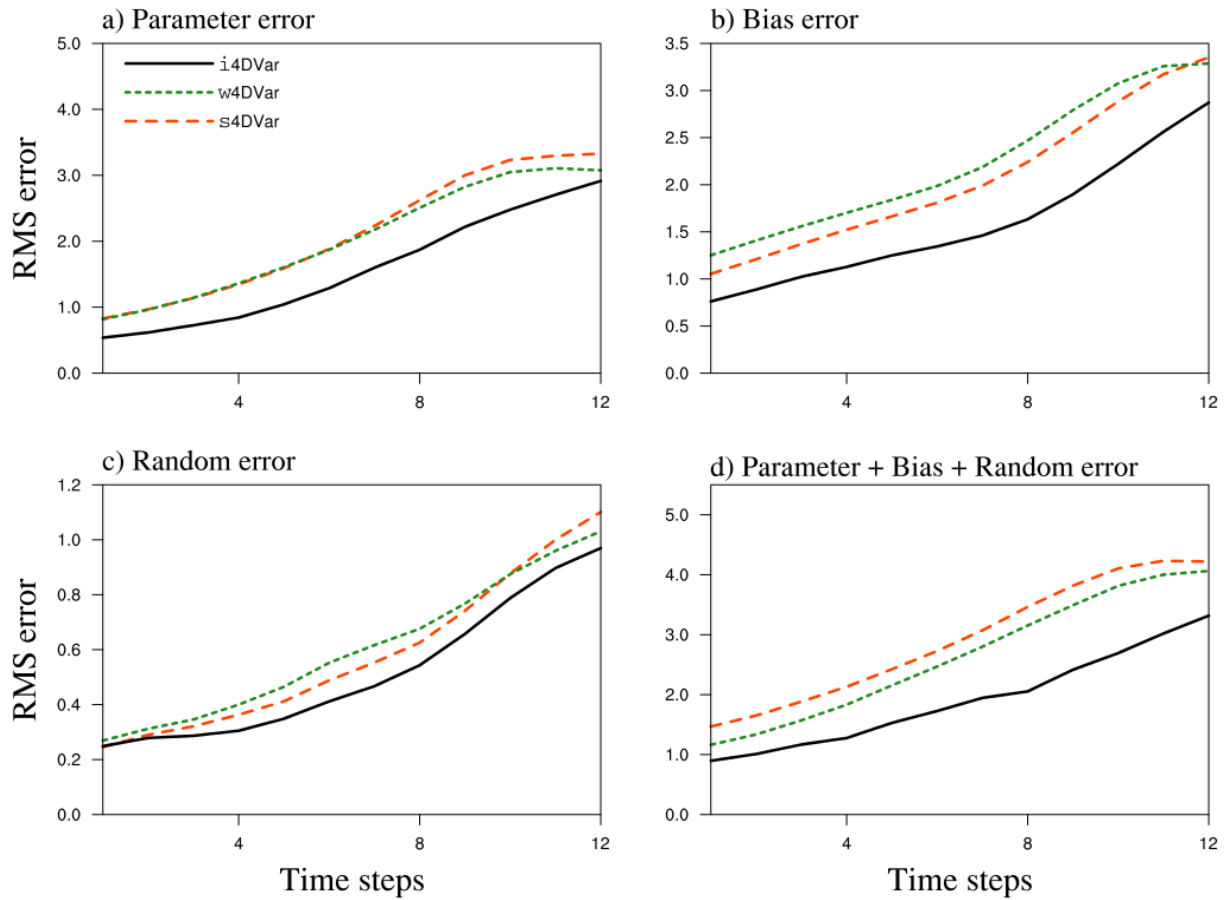

**Figure 3 | Time series of the 4DVar-driven forecast root mean square (RMS) errors using three different methods. a,** The 12-step (3-window) model forecasts after the assimilation period under the parameter error scenario. **b**, Same as **a**, but for the bias error scenario. **c**, Same as **a**, but for the random error scenario. **d**, Same as **a**, but for the parameter error + bias error + random error scenario.

**METHODS**

**Error evolution and correction within the assimilation window.** Similar to that in the w4DVar method, we consider the nonlinear model system given by

$$\mathbf{x}_k = \mathbf{f}_k(\mathbf{x}_{k-1}) + \boldsymbol{\varepsilon}_k^m, \tag{1}$$

where $\mathbf{x}_k \in \mathbb{R}^{m_x}$ is the simulated state at time step $k$, $\mathbf{f}_k : \mathbb{R}^{m_x} \to \mathbb{R}^{m_x}$ represents the nonlinear evolution of the state and $\boldsymbol{\varepsilon}_k^m \in \mathbb{R}^{m_x}$ represents model error from time step $k-1$ to $k$. For each single-step forecast, the error evolution can be formulated by a simple identity operator[20]. Thus, the integral (or total) error $\boldsymbol{\varepsilon}_k$ in $\mathbf{x}_k$ is the error $\boldsymbol{\varepsilon}_{k-1}$ (in $\mathbf{x}_{k-1}$) plus the model error $\boldsymbol{\varepsilon}_k^m$ from step $k-1$ to $k$

$$\boldsymbol{\varepsilon}_k \approx \boldsymbol{\varepsilon}_{k-1} + \boldsymbol{\varepsilon}_k^m, \ 1 \leq k \leq L_0, \tag{2}$$

where $\boldsymbol{\varepsilon}_0$ is the initial error in $\mathbf{x}_0$ at time $t_0$. If $\mathbf{x}_{k-1}$ is not corrected for each step over the assimilation window (AW), as in the s4DVar, the initial error $\boldsymbol{\varepsilon}_0$ and the model error $\boldsymbol{\varepsilon}_k^m$, will be propagated into all $\mathbf{x}_k$, and the model forecast error from earlier forecasts will accumulate with $k$, leading to increasingly large $\boldsymbol{\varepsilon}_k$ as $k$ approaches the end of the AW.

To mitigate this problem, we attempt to correct $\mathbf{x}_{k-1}$ by replacing it with $\mathbf{x}_{k-1} - \boldsymbol{\varepsilon}_{k-1}$ to remove its error $\boldsymbol{\varepsilon}_{k-1}$ before using it as the starting condition to forecast $\mathbf{x}_{k-1}$. We define $\eta_k$ such that $\boldsymbol{\varepsilon}_k^m = \eta_k \left( \boldsymbol{\varepsilon}_{k-1} + \boldsymbol{\varepsilon}_k^m \right)$ and then $\boldsymbol{\varepsilon}_{k-1} = (1 - \eta_k)(\boldsymbol{\varepsilon}_{k-1} + \boldsymbol{\varepsilon}_k^m)$. After this correction, only the model forecast error $\boldsymbol{\varepsilon}_k^m$ is passed into $\mathbf{x}_k$; so its error becomes:

$$\boldsymbol{\varepsilon}_k \approx \left( \boldsymbol{\varepsilon}_{k-1} + \boldsymbol{\varepsilon}_k^m \right) \times \eta_k. \tag{3}$$

Applying this eq. to $k-1, k-2, \cdots, 0,$ the error $\boldsymbol{\varepsilon}_k$ in $\mathbf{x}_k$ can be expressed as

$$\boldsymbol{\varepsilon}_k \approx \left( \boldsymbol{\varepsilon}_{k-1} + \boldsymbol{\varepsilon}_k^m \right) \times \eta_k$$

$$= \boldsymbol{\varepsilon}_{k-2} \times \eta_{k-1} \times \eta_k + \boldsymbol{\varepsilon}_{k-1}^m \times \eta_{k-1} \times \eta_k + \boldsymbol{\varepsilon}_k^m \times \eta_k$$

$$\cdots$$

$$= \boldsymbol{\varepsilon}_0 \times \eta_1 \times \eta_2 \times \cdots \times \eta_k + \boldsymbol{\varepsilon}_1^m \times \eta_1 \times \eta_2 \times \cdots \times \eta_k + \boldsymbol{\varepsilon}_2^m \times \eta_2 \times \eta_2 \times \cdots \times \eta_k + \cdots + \boldsymbol{\varepsilon}_k^m \times \eta_k. \quad (4)$$

To reduce the number of parameters in eq. (4), here we choose a special type of error correction in which the correction at each time removes the same faction $\upsilon$ of the error $\boldsymbol{\varepsilon}_k$ in each single-step forecast. Further, we will ignore the difference among the individual model error vectors (which is small for a fixed forecast interval within the AW), and replace them with a mean model error $\boldsymbol{\varepsilon}_{ave}^m \ (= \frac{1}{L_0} \sum_{k=1}^{L_0} \boldsymbol{\varepsilon}_k^m)$. With this approximation, eq. (4) becomes

$$\boldsymbol{\varepsilon}_k \approx \boldsymbol{\varepsilon}_0 \times \eta_1 \times \eta_2 \times \cdots \times \eta_k + \boldsymbol{\varepsilon}_1^m \times \eta_1 \times \eta_2 \times \cdots \times \eta_k + \boldsymbol{\varepsilon}_2^m \times \eta_2 \times \eta_2 \times \cdots \times \eta_k + \cdots + \boldsymbol{\varepsilon}_k^m \times \eta_k$$

$$\approx \boldsymbol{\varepsilon}_0 \upsilon^k + \boldsymbol{\varepsilon}_{ave}^m \left( \upsilon^k + \upsilon^{k-1} + \cdots + \upsilon \right)$$

$$= \boldsymbol{\varepsilon}_0 \upsilon^k + \boldsymbol{\varepsilon}_{ave}^m \frac{\upsilon - \upsilon^{k+1}}{1 - \upsilon}. \quad (5)$$

Although eq. (5) reduces the number of error vectors from $L_0 + 1$ as in w4DVar to only two, it still has one more vector than that in the s4DVar. To further simplify this error-evolution model, we introduce the maximum integral error (vector) $\boldsymbol{\varepsilon}_{max} \ (= \boldsymbol{\varepsilon}_0 + \boldsymbol{\varepsilon}_{ave}^m)$, and choose a value for $\upsilon$ such that

$$\boldsymbol{\varepsilon}_{ave}^m = \boldsymbol{\varepsilon}_{max} \upsilon, \quad (6a)$$

and

$$\boldsymbol{\varepsilon}_0 = \boldsymbol{\varepsilon}_{max}(1 - \upsilon) \quad (6b)$$

with the additional assumption that the two vectors $\boldsymbol{\varepsilon}_{ave}^m$ and $\boldsymbol{\varepsilon}_0$ have the same direction. This is a reasonable assumption given that $\boldsymbol{\varepsilon}_0$ is the error term in the assimilated state at the end of the previous AW, similar to the error term at the end of the current AW. Substituting eq. (6) into eq. (5), we obtain an evolution of $\boldsymbol{\varepsilon}_k$ as follows

$$\boldsymbol{\varepsilon}_k \approx \begin{cases} \boldsymbol{\varepsilon}_{max}(1-\upsilon), & k = 0 \\ \boldsymbol{\varepsilon}_{max} \dfrac{1}{1-\upsilon} \left[ \upsilon^2 + (1-2\upsilon)\upsilon^k \right], & 1 \leq k \leq L_0 \end{cases}. \quad (7a)$$

Since $0 < \upsilon < 1$, eq. (7) suggests that the error ($\varepsilon_k$) decreases (increases) exponentially with time index $k$ for $\upsilon < 0.5$ ($\upsilon > 0.5$). For $\upsilon < 0.5$, $\varepsilon_{ave}^m$ is less than $\varepsilon_0$, which suggests that the mean model error $\varepsilon_{ave}^m$ over one-single step (a very short time period) should be generally smaller than the initial error $\varepsilon_0$. Otherwise, the forecast model is meaningless. Particularly,

$$\varepsilon_k \approx \begin{cases} \varepsilon_{max}, & k = 0 \\ 0, & 1 \leq k \leq L_0 \end{cases} \text{ for } \upsilon = 0 \text{ and } \varepsilon_k = 0.5\varepsilon_{max} \text{ for } \upsilon = 0.5.$$ Therefore, the reasonable value range of $\upsilon$ is $[0, 0.5]$. Correspondingly, the correction term $\mathbf{x}_k'$ for the integral error $\varepsilon_k$ is thus formulated as follows

$$\mathbf{x}_k' = \begin{cases} \mathbf{x}'(1-\upsilon), & k = 0 \\ \mathbf{x}' \dfrac{1}{1-\upsilon}\left[\upsilon^2 + (1-2\upsilon)\upsilon^k\right], & 1 \leq k \leq L_0 \end{cases}, \tag{7b}$$

where $\mathbf{x}' = -\varepsilon_{max}$ is the maximum integral correction term.

We emphasize that eq. (7a) represents the evolution of only part of the total error in $\mathbf{x}_k$, since we ignored the differences among the individual model error vectors ($\varepsilon_k^m$) and set all the correction factor $\eta_k$ to the parameter $\upsilon$. In other words, the $\varepsilon_k$ from eq. (7a) represents only part (hopefully a significant part) of the total error in $\mathbf{x}_k$, and thus the adjustment of replacing $\mathbf{x}_{k-1}$ with $\mathbf{x}_{k-1} - \varepsilon_{k-1}$ also does not completely remove the error in $\mathbf{x}_{k-1}$, in contrast to what we originally attempted to do in deriving eq. (3). However, these approximations do not alter the above derivations that lead to eq. (7a); they only make the $\varepsilon_k$ part of the total error in $\mathbf{x}_k$, instead of its total error. Thus, at least for the part of the error discussed here, the structure represented by eq. (7a) is valid. Despite the approximations, our experiments with the Lorenz model suggest that the error corrections based on this simplified error model still significantly improve the assimilation and forecast results. This implies that the error model of eq. (7a) indeed represents a significant part of the total error in $\mathbf{x}_k$; otherwise, our adjustment of replacing $\mathbf{x}_{k-1}$ with $\mathbf{x}_{k-1} + \mathbf{x}_{k-1}^{'}$ (i.e. $\mathbf{x}_{k-1} - \mathbf{\varepsilon}_{k-1}$) would make little improvement.

**The ensemble nonlinear least squares-based w4DVar approach (NLS-w4DVar).** We consider one special version of the w4DVar problem[20], in which the model errors are evolving with model evolution. This w4DVar[20] seeks the analysis increment $\mathbf{x}'$ of the initial conditions and the initial model error $\boldsymbol{\varepsilon}_0^m$, such that $\mathbf{x}'$ and $\boldsymbol{\varepsilon}_0^m$ minimize the following cost function[20]:

$$\ell = \frac{1}{2}(\mathbf{x}')^{\mathrm{T}} \mathbf{B}_0^{-1}(\mathbf{x}') + \frac{1}{2}\sum_{n=0}^{S}\left[\mathbf{h}_n(\mathbf{x}_{k_n}) - \mathbf{y}_{obs,n}\right]^{\mathrm{T}} \mathbf{R}_n^{-1}\left[\mathbf{h}_n(\mathbf{x}_{k_n}) - \mathbf{y}_{obs,n}\right] + \frac{1}{2}\boldsymbol{\varepsilon}_0^{m\mathrm{T}} \mathbf{Q}_0^{-1} \boldsymbol{\varepsilon}_0^m \quad (8a)$$

subject to the forecast model

$$\mathbf{x}_k = \mathbf{f}_k(\mathbf{x}_{k-1}) + \mathbf{g}_k \boldsymbol{\varepsilon}_{k-1}^m, \quad (8b)$$

where the evolution operator $\mathbf{g}_k$ of model error is taken as the tangent linear model of the forecast model $\mathbf{f}_k$ [20]. Eq. (8) can be rewritten as follows

$$\ell = \frac{1}{2}(\mathbf{x}')^{\mathrm{T}} \mathbf{B}_0^{-1}(\mathbf{x}') + \frac{1}{2}\sum_{n=0}^{S}\left[L'_{w,k_n}(\mathbf{x}', \boldsymbol{\varepsilon}_0^m) - \mathbf{y}'_{obs,n}\right]^{\mathrm{T}} \mathbf{R}_n^{-1}\left[L'_{w,k_n}(\mathbf{x}', \boldsymbol{\varepsilon}_0^m) - \mathbf{y}'_{obs,n}\right] + \frac{1}{2}\boldsymbol{\varepsilon}_0^{m\mathrm{T}} \mathbf{Q}_0^{-1} \boldsymbol{\varepsilon}_0^m, \quad (9a)$$

subject to the forecast model

$$\mathbf{x}_k = \mathbf{f}_{w,0 \to k}(\mathbf{x}'_\#) \quad (9b)$$

where

$$\mathbf{x}'_\# = \begin{pmatrix} \mathbf{x}' \\ \boldsymbol{\varepsilon}_0^m \end{pmatrix}, \quad (10a)$$

$$\mathbf{f}_{s,0 \to k}(\mathbf{x}'_\#) = \mathbf{f}_k\left(\cdots \mathbf{f}_2\left(\mathbf{f}_1(\mathbf{x}_0) + \mathbf{g}_1 \boldsymbol{\varepsilon}_0^m\right) + \mathbf{g}_2 \mathbf{g}_1 \boldsymbol{\varepsilon}_0^m \cdots\right) + \mathbf{g}_k \cdots \mathbf{g}_2 \mathbf{g}_1 \boldsymbol{\varepsilon}_0^m, \quad (10b)$$

$$L'_{w,k_n}(\mathbf{x}'_\#) = \mathbf{h}_n \mathbf{f}_{w,0 \to k_n}(\mathbf{x}'_\#) - \mathbf{h}_n \mathbf{f}_{w,0 \to k_n}(0), \quad (10c)$$

and

$$\mathbf{y}'_{obs,n} = \mathbf{y}_{obs,n} - \mathbf{h}_n \mathbf{f}_{w,0 \to k_n}(0). \quad (10d)$$

Similar to NLS-s4DVar[24], the NLS-w4DVar approach assumes that the joint-vector $\mathbf{x}_{\#}^{'} = \begin{pmatrix} \mathbf{x}^{'} \\ \boldsymbol{\varepsilon}_0^m \end{pmatrix}$ is expressed by the linear combinations of the joint initial model perturbations (MPs) $\mathbf{x}_{\#j}^{'} = \begin{pmatrix} \mathbf{x}_j^{'} \\ \mathbf{e}_j^{'} \end{pmatrix}$ ($j=1,\cdots,N$) as follows

$$\mathbf{x}_{\#}^{'} = \mathbf{P}_x^{\#}\boldsymbol{\beta} \tag{11}$$

where $\mathbf{P}_x^{\#} = (\mathbf{x}_{\#1}^{'}, \mathbf{x}_{\#2}^{'}, \cdots, \mathbf{x}_{\#N}^{'})$ and $\boldsymbol{\beta} = (\beta_1, \beta_2, \cdots, \beta_N)^T$. Furthermore, we mark $\mathbf{P}_x = (\mathbf{x}_1^{'}, \mathbf{x}_2^{'}, \cdots, \mathbf{x}_N^{'})$ and $\mathbf{P}_e = (\mathbf{e}_1^{'}, \mathbf{e}_2^{'}, \cdots, \mathbf{e}_N^{'})$. Substituting eq.(11), and the ensemble background covariances[25] $\mathbf{B}_0 = \dfrac{(\mathbf{P}_x)(\mathbf{P}_x)^T}{N-1} \in \mathbb{R}^{m_x \times m_x}$ and $\mathbf{Q}_0 = \dfrac{(\mathbf{P}_e)(\mathbf{P}_e)^T}{N-1} \in \mathbb{R}^{m_x \times m_x}$ into eq.(9a), and expressing the cost function in terms of the new control variable $\boldsymbol{\beta}$ yield

$$\ell = \frac{1}{2} \times 2(N-1) \cdot \boldsymbol{\beta}^T \boldsymbol{\beta} + \frac{1}{2}\sum_{n=0}^{S}\left[L_{w,k_n}^{'}(\mathbf{P}_x\boldsymbol{\beta}) - \mathbf{y}_{obs,n}^{'}\right]^T \mathbf{R}_n^{-1}\left[L_{w,k_n}^{'}(\mathbf{P}_x\boldsymbol{\beta}) - \mathbf{y}_{obs,n}^{'}\right]. \tag{12}$$

After a series of mathematical transformations similar to those in formulating NLS-s4DVar[24], eq. (12) can be transformed into a non-linear least squares formulation[24,26], which is solved by the Gauss–Newton iteration scheme as follows[24,26]:

$$\boldsymbol{\beta}^i = \boldsymbol{\beta}^{i-1} - \left[2(N-1)\mathbf{I} + \sum_{n=0}^{S}(\mathbf{P}_{y,n})^T \mathbf{R}_n^{-1}(\mathbf{P}_{y,n})\right]^{-1}$$
$$\times \left[\sum_{n=0}^{S}(\mathbf{P}_{y,n})^T \mathbf{R}_n^{-1}\left(L_{w,k_n}^{'}(\mathbf{P}_x\boldsymbol{\beta}^{i-1}) - \mathbf{y}_{obs,n}^{'}\right) + 2(N-1)\boldsymbol{\beta}^{i-1}\right], \tag{13}$$

where $\mathbf{P}_{y,n} = (\mathbf{y}_{n,1}^{'}, \mathbf{y}_{n,2}^{'}, \cdots, \mathbf{y}_{n,N}^{'}) \in \mathbb{R}^{m_{y,n} \times N}$ and $\mathbf{y}_{n,j}^{'} = \mathbf{h}_n \mathbf{f}_{w,0\to k_n}(\mathbf{x}_j^{'}) - \mathbf{h}_n \mathbf{f}_{w,0\to k_n}(0)$. Therefore, we can obtain the optimized simulations over the assimilation window through $\mathbf{x}_k = \mathbf{f}_k(\mathbf{x}_{k-1}) + \mathbf{g}_k \boldsymbol{\varepsilon}_{k-1}^m$ from the analysis increment $\mathbf{x}^{'} = \mathbf{P}_x \boldsymbol{\beta}$ and the optimized initial error $\boldsymbol{\varepsilon}_0^m = \mathbf{P}_e \boldsymbol{\beta}$.

**The ensemble nonlinear least squares-based i4DVar approach (NLS-i4DVar).** The i4DVar seeks the maximum correction-term $\mathbf{x}^{'}$ that minimizes the following cost function

$$\ell = \frac{1}{2}\left(\mathbf{x}'\right)^{\mathrm{T}} \mathbf{Q}_b^{-1}\left(\mathbf{x}'\right) + \frac{1}{2}\sum_{n=0}^{S}\left[\mathbf{h}_n(\mathbf{x}_{k_n}) - \mathbf{y}_{obs,n}\right]^{\mathrm{T}} \mathbf{R}_n^{-1}\left[\mathbf{h}_n(\mathbf{x}_{k_n}) - \mathbf{y}_{obs,n}\right] \tag{14a}$$

subject to the forecast model with the parameter $\upsilon \in [0, 0.5]$

$$\mathbf{x}_k = \begin{cases} \mathbf{f}_k\left(\mathbf{x}_{k-1} + \mathbf{x}'(1-\upsilon)\right), & k = 1 \\ \mathbf{f}_k\left(\mathbf{x}_{k-1} + \mathbf{x}' \frac{1}{1-\upsilon}\left[\upsilon^2 + (1-2\upsilon)\upsilon^{k-1}\right]\right), & 2 \le k \le L_0 \end{cases}. \tag{14b}$$

Similarly, eq. (14) can be also rewritten into the following format

$$\ell = \frac{1}{2}\left(\mathbf{x}'\right)^{\mathrm{T}} \mathbf{Q}_b^{-1}\left(\mathbf{x}'\right) + \frac{1}{2}\sum_{n=0}^{S}\left[L_{k_n}^{\upsilon'}(\mathbf{x}') - \mathbf{y}'_{obs,n}\right]^{\mathrm{T}} \mathbf{R}_n^{-1}\left[L_{k_n}^{\upsilon'}(\mathbf{x}') - \mathbf{y}'_{obs,n}\right] \tag{15a}$$

subject to the forecast model

$$\mathbf{x}_k = \mathbf{f}_{i,0\to k}^{\upsilon}(\mathbf{x}'), \tag{15b}$$

where

$$\mathbf{f}_{i,0\to k}^{\upsilon}(\mathbf{x}') = \mathbf{f}_k\left(\mathbf{f}_{k-1}\left(\cdots \mathbf{f}_2\left(\mathbf{f}_1\left(\mathbf{x}_0 + \mathbf{x}'(1-\upsilon)\right) + \mathbf{x}' \frac{1}{1-\upsilon}\left[\upsilon^2 + (1-2\upsilon)\upsilon\right]\right)\cdots\right)\right.$$

$$\left. + \mathbf{x}' \frac{1}{1-\upsilon}\left[\upsilon^2 + (1-2\upsilon)\upsilon^{k-1}\right]\right), \tag{16a}$$

$$L_{k_n}^{\upsilon'}(\mathbf{x}') = \mathbf{h}_n \mathbf{f}_{i,0\to k_n}^{\upsilon}(\mathbf{x}') - \mathbf{h}_n \mathbf{f}_{i,0\to k_n}^{\upsilon}(0), \tag{16b}$$

and

$$\mathbf{y}'_{obs,n} = \mathbf{y}_{obs,n} - \mathbf{h}_n \mathbf{f}_{i,0\to k_n}^{\upsilon}(0). \tag{16c}$$

Similarly, the maximum integral correction-term $\mathbf{x}'$ is assumed to be expressed by the linear combinations of the MPs ($\mathbf{x}'_j$, $j = 1, \cdots, N$) as follows

$$\mathbf{x}' = \mathbf{P}_x \boldsymbol{\beta} \tag{17}$$

where $\mathbf{P}_x = (\mathbf{x}_1', \mathbf{x}_2', \cdots, \mathbf{x}_N')$ and $\boldsymbol{\beta} = (\beta_1, \beta_2, \cdots, \beta_N)^{\mathrm{T}}$. Substituting eq. (17) and the ensemble error covariance[25] $\mathbf{Q}_b = \dfrac{(\mathbf{P}_x)(\mathbf{P}_x)^{\mathrm{T}}}{N-1} \in \mathbb{R}^{m_x \times m_x}$ into eq. (15a) and expressing the cost function in terms of the new control variable $\boldsymbol{\beta}$ yield

$$\ell = \frac{1}{2}(N-1) \cdot \boldsymbol{\beta}^{\mathrm{T}} \boldsymbol{\beta} + \frac{1}{2} \sum_{n=0}^{S} \left[ L_{k_n}^{\upsilon'}(\mathbf{P}_x \boldsymbol{\beta}) - \mathbf{y}_{obs,n}' \right]^{\mathrm{T}} \mathbf{R}_n^{-1} \left[ L_{k_n}^{\upsilon'}(\mathbf{P}_x \boldsymbol{\beta}) - \mathbf{y}_{obs,n}' \right]. \tag{18}$$

Similarly, after a series of mathematical transformations similar to those in formulating NLS-s4DVar[24], eq.(15a) can be also transformed into a non-linear least squares formulation[24,26], which is solved by the Gauss–Newton iteration scheme as follows[24,26]:

$$\boldsymbol{\beta}^i = \boldsymbol{\beta}^{i-1} - \left[ (N-1)\mathbf{I} + \sum_{n=0}^{S} (\mathbf{P}_{y,n})^{\mathrm{T}} \mathbf{R}_n^{-1} (\mathbf{P}_{y,n}) \right]^{-1}$$
$$\times \left[ \sum_{n=0}^{S} (\mathbf{P}_{y,n})^{\mathrm{T}} \mathbf{R}_n^{-1} \left( L_{k_n}^{\upsilon'}(\mathbf{P}_x \boldsymbol{\beta}^{i-1}) - \mathbf{y}_{obs,n}' \right) + (N-1)\boldsymbol{\beta}^{i-1} \right], \tag{19}$$

where $\mathbf{P}_{y,n} = (\mathbf{y}_{n,1}', \mathbf{y}_{n,2}', \cdots, \mathbf{y}_{n,N}') \in \mathbb{R}^{m_{y,n} \times N}$ and $\mathbf{y}_{n,j}' = \mathbf{h}_n \mathbf{f}_{i,0 \to k_n}(\mathbf{x}_j') - \mathbf{h}_n \mathbf{f}_{i,0 \to k_n}(0)$. As a result, we can obtain the optimized model simulations over the assimilation window through $\mathbf{f}_k \left( \mathbf{x}_{k-1} + \mathbf{x}' \dfrac{1}{1-\upsilon} \left[ \upsilon^2 + (1-2\upsilon)\upsilon^k \right] \right)$ from the optimized maximum correction-term $\mathbf{x}' = \mathbf{P}_x \boldsymbol{\beta}$ with the predetermined parameter $\upsilon$.

**Localization Implementation of the NLS-w/i4DVar approaches.** An important issue in the ensemble-based method is sampling errors, and a practical way to address this issue is through a localization technique, which could ameliorate the contaminations resulting from inadequate sampling or the spurious long-range correlations[27]. Obviously, the final Gauss–Newton iteration schemes to the i4DVar and w4DVar problems share the following unified formulation

$$\boldsymbol{\beta}^i = \boldsymbol{\beta}^{i-1} - \left[ \gamma(N-1)\mathbf{I} + \sum_{n=0}^{S} (\mathbf{P}_{y,n})^{\mathrm{T}} \mathbf{R}_n^{-1} (\mathbf{P}_{y,n}) \right]^{-1}$$
$$\times \left[ \sum_{n=0}^{S} (\mathbf{P}_{y,n})^{\mathrm{T}} \mathbf{R}_n^{-1} \left( L_{k_n}^{'(\gamma)}(\mathbf{P}_x \boldsymbol{\beta}^{i-1}) - \mathbf{y}_{obs,n}' \right) + \gamma(N-1)\boldsymbol{\beta}^{i-1} \right], \tag{20}$$

where $\gamma=1$, $L_{k_n}^{'(1)}=L_{k_n}^{\upsilon'}$ for NLS-i4DVar, and $\gamma=2$, $L_{k_n}^{'(2)}=L_{w,k_n}^{'}$ for NLS-w4DVar, respectively.

We adapt an efficient localization scheme proposed for the NLS-s4DVar[24] to transform eq. (20) into the following form:

$$\boldsymbol{\beta}^i = \boldsymbol{\beta}^{i-1} + \sum_{n=0}^{S}\left[\left(\mathbf{P}_{y,n}^{\#} <e> \boldsymbol{\rho}_{y,n}\right)^{\mathrm{T}} L_{k_n}^{'(\gamma)}(\mathbf{P}_{x,\rho}\boldsymbol{\beta}^{i-1})\right]$$

$$+\sum_{n=0}^{S}\left(\mathbf{P}_{y,n}^{*} <e> \boldsymbol{\rho}_{y,n}\right)^{\mathrm{T}} \mathbf{R}_n^{-1}\left[\mathbf{y}_{obs,n}^{'} - L_{k_n}^{'(\gamma)}(\mathbf{P}_{x,\rho}\boldsymbol{\beta}^{i})\right], \quad (21a)$$

where

$$\mathbf{P}_{y,n}^{\#\ \mathrm{T}} = -\gamma(N-1)\left[\sum_{n=0}^{S}\left(\mathbf{P}_{y,n}\right)^{\mathrm{T}}\mathbf{R}_n^{-1}\left(\mathbf{P}_{y,n}\right)+\gamma(N-1)\mathbf{I}\right]^{-1}\left[\sum_{n=0}^{S}\left(\mathbf{P}_{y,n}\right)^{\mathrm{T}}\left(\mathbf{P}_{y,n}\right)\right]^{-1}\left(\mathbf{P}_{y,n}\right)^{\mathrm{T}}, \quad (21b)$$

and

$$\mathbf{P}_{y,n}^{*\ \mathrm{T}} = \left[\sum_{n=0}^{S}\left(\mathbf{P}_{y,n}\right)^{\mathrm{T}}\mathbf{R}_n^{-1}\left(\mathbf{P}_{y,n}\right)+\gamma(N-1)\mathbf{I}\right]^{-1}\left(\mathbf{P}_{y,n}\right)^{\mathrm{T}}. \quad (21c)$$

Then we can obtain $\mathbf{x}^{',i} = \mathbf{P}_{x,\rho}\boldsymbol{\beta}^i$, where

$$\mathbf{P}_{x,\rho} = \left(\mathbf{P}_x <e> \boldsymbol{\rho}_x\right) = \left(\boldsymbol{\rho}_x \circ \mathbf{P}_{x,1}^{*}, \cdots, \boldsymbol{\rho}_x \circ \mathbf{P}_{x,N}^{*}\right) \in \mathbb{R}^{m_x \times (r \times N)} \quad (22)$$

($r$ is the is the eigenvector number chosen[28]) is the localized MP matrix, and $\mathbf{P}_{x,j}^{*}$ ($j=1,\cdots,N$) is an $m_x \times r$ matrix whose every column is the $j$th column of $\mathbf{P}_x$, $\boldsymbol{\rho}_{y,n} \in \mathbb{R}^{m_{y,n} \times r}$, $\boldsymbol{\rho}_x \in \mathbb{R}^{m_x \times r}$, and $\boldsymbol{\rho}_n = \boldsymbol{\rho}_x \boldsymbol{\rho}_{y,n}^{\mathrm{T}}$. Here, $\boldsymbol{\rho}_n \in \mathbb{R}^{m_x \times m_{y,n}}$ is the spatial correlation matrix[28], and $\mathbf{B} \circ \mathbf{C}$ stands for the Schür product of matrices and $\mathbf{C}$ which is a matrix whose $(i,j)$ entry is given by $b_{i,j} \times c_{i,j}$. Eq. (22) also defines the matrix operator $(\bullet <e> \bullet)$ for two matrices.

**Model and experimental Design.** We use the Lorenz-96 model[23] as the model platform to evaluate all the three 4DVar methods. This model has been widely used to study various issues associated with data assimilation. The Lorenz-96 model is governed by the following system of nonlinear equations:

$$\frac{dx_i}{dt} = -x_{i-2}x_{i-1} + x_{i-1}x_{i+1} - x_i + F, i = 1, \cdots, n, \tag{23}$$

with periodic boundary conditions. The model behaves quite differently with different values of $F$ and produces chaotic systems with integer values of $F$ larger than three. In this configuration, we take $n = 40$ and $F = 8$. For computational stability, a time step of 0.05 units (or 6 h in equivalent[23]) is adopted and a fourth-order Runge–Kutta scheme is used for temporal integration in this study.

An observing system simulation experiment (OSSE) is considered as one of the best benchmark tests to evaluate a data assimilation methodology since it can provide both the 'true' state and the corresponding 'observation'. These artificial observations were assimilated in the OSSEs: The default number of observations is $m = 20$ (equally spaced at every observation time) and observations were taken every two time steps (or 12 h), which were generated by adding random error perturbations with the standard deviation error of 0.1 to the time series of the true state obtained by running the perfect model ($F = 8$) from the "true" initial field. All experiments were carried out over 6 days. The default parameter setups are the ensemble size $N = 30$, $r = 10$, and the covariance localization Schür radius $r_0 = d_{h,0}(d_{v0}) = 16$ (only one direction in the Lorenz-96 model). The length of assimilation window is 4 steps and the ensemble is re-generated by the LETKF format[29] for each cycle. The background initial field is generated by integrating the model 100000 steps from an arbitrary nonzero fields and the true initial field is subsequently obtained by integrating the 40 steps from the background initial field. Therefore, this background state is significantly different from the "true" state. The initial sample is generated by running the Lorenz model from 90 steps (i.e., sampling once every three time steps). To verify the three 4DVar approaches comprehensively, we design one group of model error scenarios (totally four cases) including the parameter error ($F = 11$), constant bias error $\mathbf{v}_s$ ($\|\mathbf{v}_s\|=1.5$), random error $\mathbf{v}_r \in N(0,0.1)$ and their combinations (parameter error + bias error + random error).

**Availability.** All other relevant data and codes are available from the corresponding author.

# Extended Data

**Extended Data Table 1. | Summary of key notation used in this article**

| Notation | Description | Size |
|---|---|---|
| $\mathbf{x}_0, \mathbf{x}', \mathbf{x}'_j$ | State vector: time $t_0$ ( $'$ increment, $j$ ensemble member) | $m_x$ |
| $\mathbf{x}_{k-1}, \mathbf{x}^*_{k-1}$ | State vector: time step $k-1$ ($^*$ corrected) | $m_x$ |
| $\mathbf{B}, \mathbf{B}_0, \mathbf{Q}_b$ | Background error covariance | $m_x \times m_x$ |
| $\mathbf{R}_n$ | Observation error covariance | $m_{y,n} \times m_{y,n}$ |
| $\mathbf{f}_k(\cdot)$ | Forecast model: time step $k-1 \to k$ | $m_x$ (in, out) |
| $\mathbf{h}_n(\cdot)$ | Nonlinear observation operator: time $t_n$ | $m_x$ (in), $m_{y,n}$ (out) |
| $\mathbf{y}_{obs,n}$ | Observations: time $t_n$ | $m_{y,n}$ |
| $\boldsymbol{\varepsilon}^m_0, \boldsymbol{\varepsilon}^m_k$ | Model error terms: time step $0$, $k$ | $m_x$ |
| $\mathbf{Q}_0, \mathbf{Q}_k$ | Model error covariance | $m_x \times m_x$ |
| $\upsilon$ | Decaying parameter: $[0, 0.5]$ | 1 |
| $\mathbf{f}_{w,0\to k_n}(\cdot), \mathbf{f}^\upsilon_{i,0\to k_n}(\cdot)$ | Forecast model: time $t_0 \to t_n$ ($w$ weak, $i$ integral) | $m_x$ (in, out) |
| $L'_{w,k_n}(\mathbf{x}')$ | $L'_{w,k_n}(\cdot) = \mathbf{h}_n\mathbf{f}_{w,0\to k_n}(\cdot) - \mathbf{h}_n\mathbf{f}_{w,0\to k_n}(0)$ : time $t_0 \to t_n$ | $m_x$ (in), $m_{y,n}$ (out) |
| $L^{\upsilon'}_{k_n}(\mathbf{x}')$ | $L^{\upsilon'}_{k_n}(\cdot) = \mathbf{h}_n\mathbf{f}^\upsilon_{i,0\to k_n}(\cdot) - \mathbf{h}_n\mathbf{f}^\upsilon_{i,0\to k_n}(0)$ : time $t_0 \to t_n$ | $m_x$ (in), $m_{y,n}$ (out) |
| $\mathbf{y}'_{obs,n}$ | $\mathbf{y}_{obs,n} - \mathbf{h}_n\mathbf{f}_{w,0\to k_n}(0)$ ; $\mathbf{y}_{obs,n} - \mathbf{h}_n\mathbf{f}^\upsilon_{i,0\to k_n}(0)$ | $m_{y,n}$ |
| $\mathbf{P}_x$ | $(\mathbf{x}'_1, \mathbf{x}'_2, \cdots, \mathbf{x}'_N)$ | $m_x \times N$ |
| $\boldsymbol{\beta}^i$ | Coefficient vector at $i$th iteration | $r \times N$ |
| $\mathbf{P}_{y,n}$ | $(\mathbf{y}'_{n,1}, \cdots, \mathbf{y}'_{n,N})$ | $m_{y,n} \times N$ |
| $\mathbf{y}'_{n,j}$ | $\mathbf{h}_n\mathbf{f}_{w,0\to k_n}(\mathbf{x}'_j) - \mathbf{h}_n\mathbf{f}_{w,0\to k_n}(0)$ ; $\mathbf{h}_n\mathbf{f}^\upsilon_{i,0\to k_n}(\mathbf{x}'_j) - \mathbf{h}_n\mathbf{f}^\upsilon_{i,0\to k_n}(0)$ | $m_{y,n}$ |
| $\mathbf{g}_k$ | The evolution operator of model error | $m_x$ (in, out) |
| $\mathbf{x}'_\#$ | Joint-vector | $2m_x$ |
| $\mathbf{P}^\#_x$ | $(\mathbf{x}'_{\#1}, \mathbf{x}'_{\#2}, \cdots, \mathbf{x}'_{\#N})$ | $2m_x \times N$ |
| $\mathbf{P}_e$ | $(\mathbf{e}'_1, \mathbf{e}'_2, \cdots, \mathbf{e}'_N)$ | $m_x \times N$ |
| $\boldsymbol{\rho}_n$ | Spatial correlation matrix | $m_x \times m_{y,n}$ |
| $\boldsymbol{\rho}_x, \boldsymbol{\rho}_{y,n}$ | $\boldsymbol{\rho}_x \boldsymbol{\rho}^T_{y,n} = \boldsymbol{\rho}_n$ | $m_x \times r$, $m_{y,n} \times r$ |
| $\mathbf{P}_{x,\rho}$ | $\mathbf{P}_x <e> \boldsymbol{\rho}_x$ | $m_x \times (r \times N)$ |

$m_x$ is the size of the state space.

$m_{y,n}$ is the size of observations at $t_n$.

$S+1$ is the total number of observation time points in the assimilation window.

$L_0$ is the total length of the assimilation window.

$r$ is the number of eigenvectors chosen in the SVD decomposition of correlation matrix $\boldsymbol{\rho}_n$.

$N$ is the number of ensemble members.

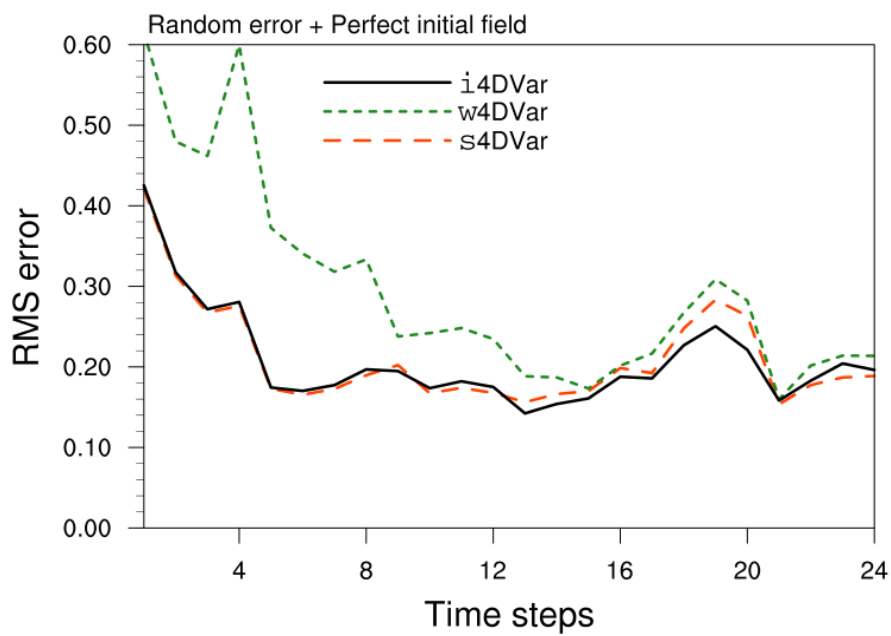

**Extended Data Figure 1 | Time series of the 4DVar-assimilated root mean square (RMS) errors.** Shown are the 24-step (6-window) assimilation results under the random error scenario with the perfect initial field.

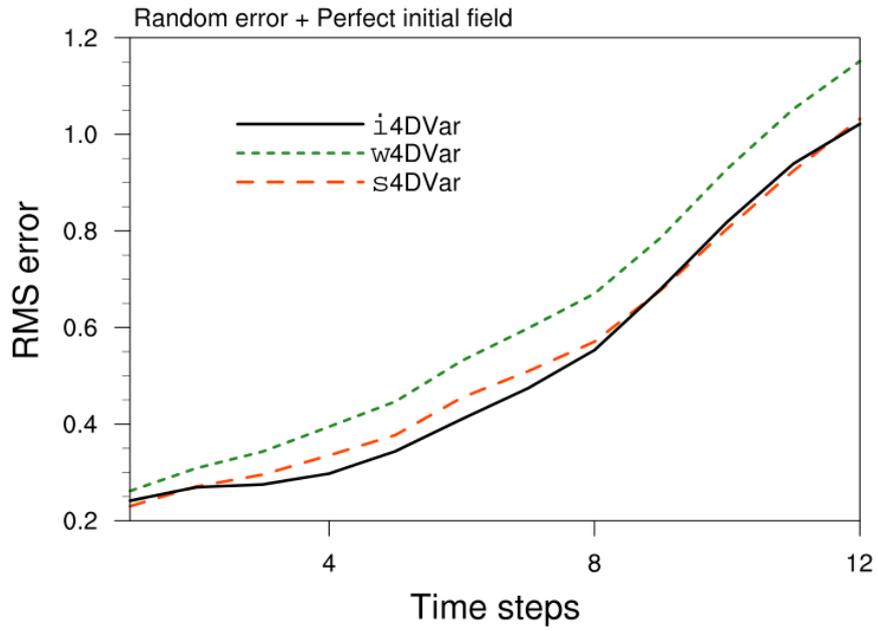

**Extended Data Figure 2 | Time series of the 4DVar-driven forecast root mean square (RMS) errors.** Shown are the 12-step (3-window) model forecasts after the assimilation period under the random error scenario with the perfect initial field.

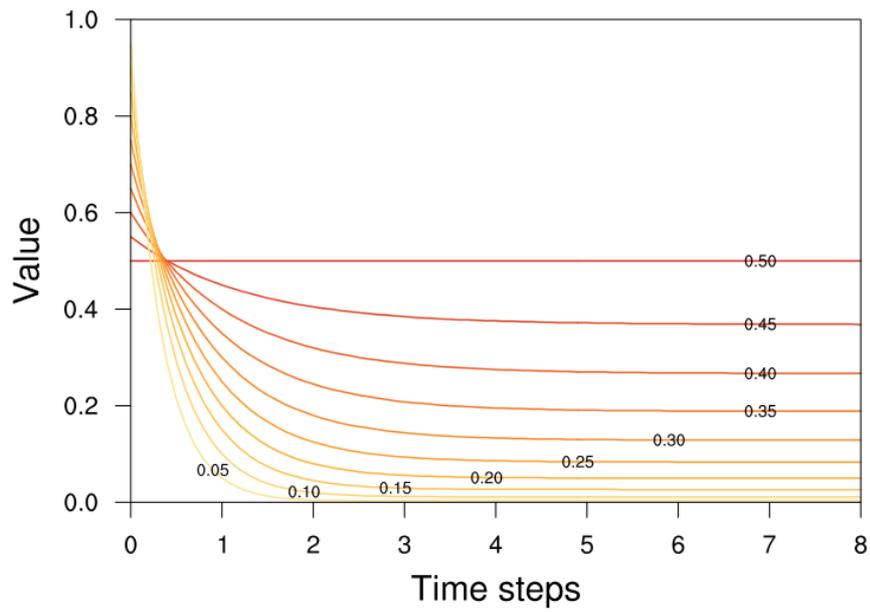

**Extended Data Figure 3 | The exponential function** $\frac{1}{1-\upsilon}\left[\upsilon^2+(1-2\upsilon)\upsilon^k\right]$. Shown are its values vary with the time step ($k$) for different $\upsilon$ values: $\upsilon = 0.5 - 0.05 \times (i-1)$, with $i = 1, \cdots, 10$.

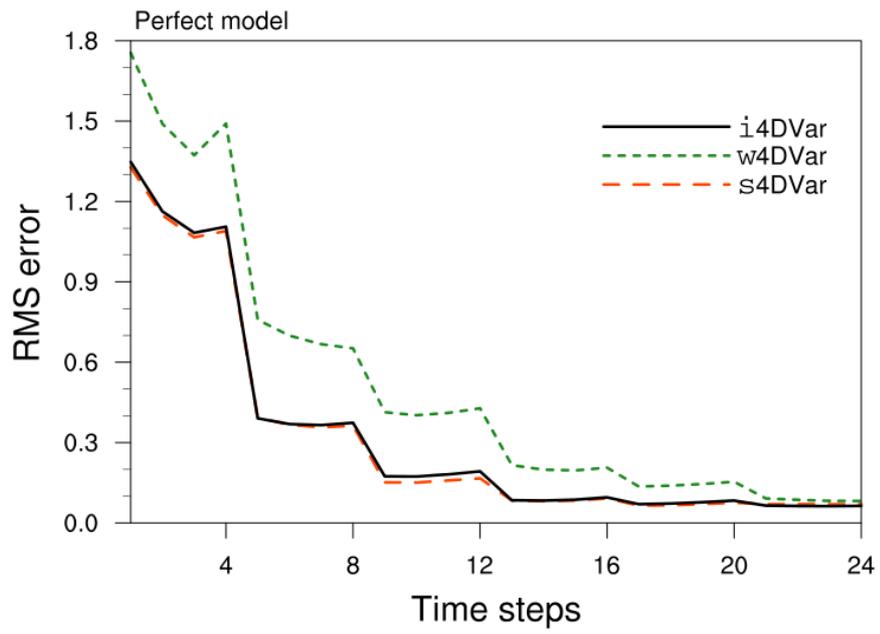

**Extended Data Figure 4 |** Same as Extended Data Fig. 1, but for the perfect model scenario with the background field.

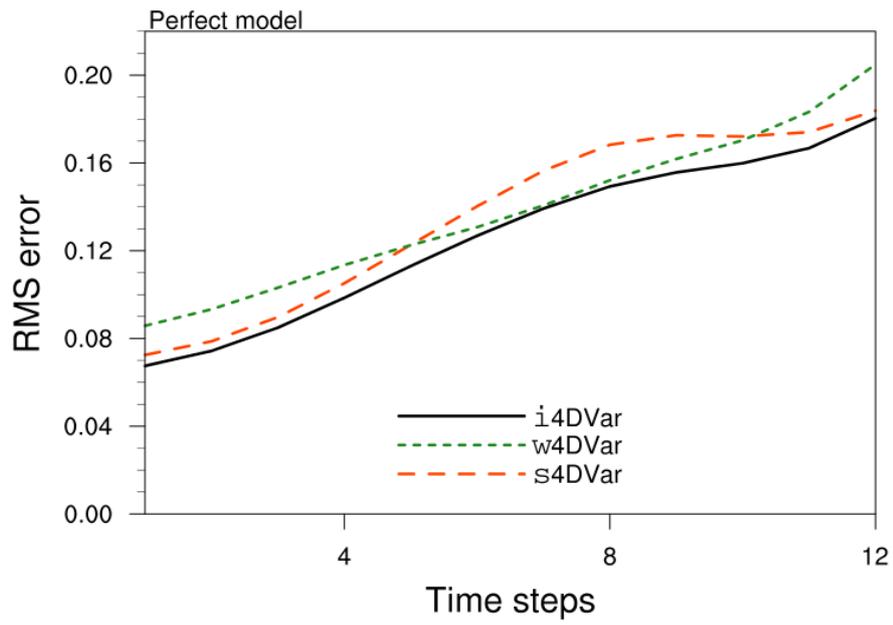

**Extended Data Figure 5 |** Same as Extended Data Fig. 2, but for the perfect model scenario with the background field.